\documentclass[11pt]{article}
\usepackage{amsmath,amssymb,color,epsfig,cite}
\usepackage{graphicx}
\usepackage{subfigure}
\usepackage{setspace}

\usepackage{amsfonts}

\newcommand{\hoch}[1]{$\, ^{#1}$}

\makeatletter
\@addtoreset{equation}{section}
\makeatother

\newcommand{\be}{\begin{equation}}
\newcommand{\ee}{\end{equation}}
\newcommand{\bea}{\setlength\arraycolsep{2pt} \begin{eqnarray}}
\newcommand{\eea}{\end{eqnarray}}
\newcommand{\nn}{\nonumber}

\def\ft#1#2{{\textstyle{\frac{\scriptstyle #1}{\scriptstyle #2} } }}
\def\fft#1#2{{\frac{#1}{#2}}}

\def\0{{\sst{(0)}}}
\def\1{{\sst{(1)}}}
\def\2{{\sst{(2)}}}
\def\3{{\sst{(3)}}}
\def\4{{\sst{(4)}}}
\def\5{{\sst{(5)}}}
\def\6{{\sst{(6)}}}
\def\7{{\sst{(7)}}}
\def\8{{\sst{(8)}}}
\def\sst#1{{\scriptscriptstyle #1}}

\begin{document}


\begin{center}
{\large {\bf Black Holes, Dark Wormholes and Solitons  in $f(T)$ Gravities}}

\vspace{10pt}
Zhan-Feng Mai and  H. L\"u\hoch{*}

\vspace{10pt}

{\it Department of Physics, Beijing Normal University,
Beijing 100875, China}

\vspace{40pt}

\underline{ABSTRACT}

\end{center}

By choosing an appropriate vielbein basis, we obtain a class of spherically-symmetric solutions in $f(T)$ gravities.  The solutions are asymptotic to Minkowski spacetimes with leading falloffs the same as those of the Schwarzschild black hole. In general, these solutions have branch-cut singularities in the middle.  For appropriately chosen $f(T)$ functions, extremal black holes can also emerge. Furthermore, we obtain wormhole configurations whose spatial section is analogous to an Ellis wormhole, but $-g_{tt}$ runs from 0 to $1$ as the proper radial coordinate runs from $-\infty$ to $+\infty$. Thus a signal sent from $-\infty$ to $+\infty$ through the wormhole will be infinitely red-shifted.  We call such a spacetime configuration a dark wormhole.  By introducing a bare cosmological constant $\Lambda_0$, we construct smooth solitons that are asymptotic to local AdS with an effective $\Lambda_{\rm eff}$.  In the middle of bulk, the soliton metric behaves like the AdS of bare $\Lambda_0$ in global coordinates. We also embed AdS planar and Lifshitz black holes in $f(T)$ gravities.  Finally we couple the Maxwell field to the $f(T)$ theories and construct electrically-charged solutions.

\vfill {\footnotesize  \hoch*mrhonglu@gmail.com}

\thispagestyle{empty}

\pagebreak

\tableofcontents
\addtocontents{toc}{\protect\setcounter{tocdepth}{2}}


\newpage

\section{Introduction}

In the original formulation of Einstein's theory of gravity, the metric is treated as the fundamental field, and the resulting theory is of the second-order in differentiations. There however can exist other formalisms.  The fact that Riemann tensor depends only on the connection leads naturally to the Palatini formalism, which may in fact have been invented by Einstein himself.  Another intriguing formulation is the teleparallel equivalent of general relativity (TEGR), which was also originated by Einstein to unify gravitation and electromagnetism \cite{Unzicker:2005in} (See also,\cite{deAndrade:1997gka,Maluf:2013gaa}.) As two-derivative theories, these different formulations of gravity are classically equivalent.  However, they may become inequivalent when higher-order terms are introduced.

In this paper, we study $f(T)$ gravity which is a generalization of TEGR, analogous to the $f(R)$ generalization of the Einstein-Hilbert action.  We focus on the construction of static solutions, with spherical or toroidal isometries.  In TEGR , it is the vielbein that is treated as fundamental fiends.  While $f(T)$ gravity is invariant under the general coordinate transformations, it is not invariant under the local Lorentz transformation of the tangent spacetime \cite{Li:2010cg}.  This implies that for a given metric ansatz, there are six-parameter worth of inequivalent vielbein choices.

One application of $f(T)$ gravity is to study early cosmology. (See for examples \cite{Bengochea:2008gz,Linder:2010py,Cai:2015emx}.) In this case, there is, {\it a priori}, no restriction that the theory should reproduce Einstein's gravity.  The purpose of this paper is to study static solutions of $f(T)$ gravity. The success of Einstein's theory in explaining the solar system puts a severe restriction of any modified gravity. In this paper, we are largely concerned with $f(T)$ gravity with
\be
f(T)=-T + \alpha T^n\,.\label{FT1}
\ee
The theory can be viewed as deviation away from Einstein gravity and it reduces to Einstein gravity when $\alpha=0$.

One particularly important class of solutions in any theory of gravity are those with spherical isometries. In general relativity, static and spherically-symmetric ansatz leads to the celebrated Schwarzschild black hole. Its most convenient vielbein base in the framework of Einstein gravity is the diagonal one; however, such an ansatz leads to no solution in a generic $f(T)$ theory.  Cleverer ansatz of the vielbein have to be considered in order to construct spherically-symmetric solutions in $f(T)$ theories. For example, general static and spherically symmetric ansatz with three parameters of the local $SO(3)\subset SO(1,3)$ were considered in \cite{Tamanini:2012hg}, which yielded the Schwarzschild-AdS metric in global coordinates. Indeed, a variety of similar ansatz were proposed in literature \cite{Nashed:uja,Boehmer:2011gw,Nashed:2014sea,Atazadeh:2012am,
Rodrigues:2013ifa,Capozziello:2012zj,Nashed:2016nce,Li:2016auv}.  In particular, (charged) Schwarzschild (anti-de Sitter (AdS)) black holes were constructed in $f(T)$ gravities\cite{Wang:2011xf,Nashed:2013bfa,Iorio:2016sqy}. (See also\cite{Nashed:2013hba,Gonzalez:2014pwa,Gonzalez:2011dr,
Hanafy:2015yya,Daouda:2011rt,Yang:2012hu}.) Interestingly, Kerr solutions were also embedded in $f(T)$ gravities\cite{Nashed:2016ywo,Nashed:2016qbe,Bejarano:2014bca}.

In this paper, one of our goal is to construct static and spherically-symmetric solutions. We regard the Minkowski spacetime in Cartesian coordinate with the diagonal vielbein basis as the most symmetric vacuum in an $f(T)$ theory.  We are thus interested in finding solutions that are asymptotic to this vacuum. We derive the vielbein ansatz that guarantees this requirement.  We study how the asymptotic behavior of the Schwarzschild black hole is modified by the $\alpha$ term in (\ref{FT1}). This behavior is important for any modified gravity to pass the solar-system tests. We are also interested in whether new and more exotic solutions such as wormholes or solitons can arise in $f(T)$ gravities within our vielbein ansatz.

Another focus of our paper is to construct asymptotically AdS solutions.  Whilst the AdS spacetime in global coordinates are subtler to construct, the planar AdS spacetime arises naturally in $f(T)$ gravity using the diagonal vielbein basis.  We construct Reissner-Nordstr\o m (RN) planar AdS black holes in $f(T)$ theories.

The paper is organized as follows.  In section 2, we give a quick review of $f(T)$ gravities and present the corresponding covariant equations of motion.  In section 3, we discuss the ansatz of vielbein. Instead of considering the most general ansatz with the six $SO(1,3)$ parameters, we focus only on those that regards the Minkowski spacetime with diagonal vielbein as the preferred frame.  In sections 4, 5, 6 and 7, we construct a variety of solutions in $f(T)$ gravities.  We conclude our paper in section 8.

\section{$f(T)$ gravities and covariant equations}

In this section we study $f(T)$ gravities. We shall present only the bare minimum that are relevant for constructing the Lagrangian and deriving covariant equations of motion. For a more complete review of teleparallel gravity, see, e.g.~\cite{Aldrovandi:2013wha}.

In Einstein's general relativity, the metric $ds^2=g_{\mu\nu} dx^\mu dx^\nu$ is the fundamental field. The curvature tensors, including the Riemann tensor $\bar R^{\mu}{}_{\nu\rho\sigma}$, Ricci tensor $\bar R_{\mu\nu}$ and Ricci scalar $\bar R$, can be constructed from the Levita-Civita connection of vanishing torsion
\begin{equation}
\Gamma^\rho{}_{\mu\nu}=\frac{1}{2}g^{\rho\lambda}(\partial_\mu g_{\lambda\nu}+\partial_\nu g_{\lambda\mu}-\partial_\lambda g_{\mu\nu})\,.
\end{equation}
The fundamental field in $f(T)$ gravity, on the other hand, is the vielbein $e^a =e^a_\mu dx^\mu$, defined by
\begin{equation}
g_{\mu\nu}=e^a_\mu e^b_\nu\, \eta_{ab}\,.
\end{equation}
Note that we use Greek letters for spacetime indices and Latin letters for indices in local Minkowski tangent space of the signature $\eta_{ab} = {\rm diag}\{-1,1,1,1\}$. It is also convenient to introduce the inverse vielbein $e^{\mu}_a$, satisfying
\be
e^a_\mu e^\mu_b=\delta^a_b\,,\qquad
e^\mu_a e^a_\nu=\delta^\mu_\nu\,.
\ee
In the framework of the teleparallel equivalent of general relativity (TEGR), for vanishing spin connections, one can define the torsion tensor:
\begin{equation}
T^a{}_{\mu\nu}=\partial_\mu e^a_\nu-\partial_\nu e^a_\mu\,,\qquad
T^\lambda{}_{\mu\nu}=e^\lambda_a T^a{}_{\mu\nu}\,.\label{torsiondef}
\end{equation}
It is clear that the torsion tensor is covariant under the general coordinate transformation, but not under the local Lorentz transformation on the tangent flat indices. The torsion vector follows straightforwardly as
\begin{equation}
T_\mu=T^\lambda{}_ {\lambda\mu}\,.
\end{equation}
The proper scalar quantity for the torsion is more subtle to define. It is convenient first to introduce contorsion tensor $K^\lambda{}_{\mu\nu}$, which is related to the torsion tensor as
\begin{equation}
K^{\rho\mu\nu}=\ft12(T^{\mu\rho\nu}+T^{\nu\rho\mu}-T^{\rho\mu\nu})\,.
\end{equation}
A superpotential tensor can be constructed as a linear combination of contorsion and torsion vector
\begin{equation}
S^{\rho\mu\nu}=\ft{1}{2} \big(K^{\mu\nu\rho} +g^{\rho\nu}T^{\mu}-g^{\rho\mu}T^{\nu}\big)\,.
\end{equation}
With these preliminaries, one can define the torsion scalar
\begin{equation}
T=S^{\rho\mu\nu}T_{\rho\mu\nu}\,.
\end{equation}
It turns out that the torsion scalar is related to the Ricci scalar $\bar R$ mentioned earlier as
\begin{equation}
T=-\bar{R}+2\nabla^{\nu}T_{\nu}\,.
\end{equation}
It is remarkable to note that $T$ has no second derivatives, and the second derivatives in $\bar R$ are all cancelled by the divergence of the torsion vector.  This implies that if we use $T$ to construct a Lagrangian, the Euler-Lagrange equation arises from the variation principle straightforwardly without needing any surface terms.

Indeed one can construct $f(T)$ gravity whose Lagrangian density is a generic function the torsion scalar.  Assuming that the matter fields are minimally coupled to the metric, we have the Lagrangian
\be
S=\fft{1}{16\pi} \int d^4x\, e\,\big(f(T)+8\pi L_{\rm min}\big)\,.
\ee
where $e=\sqrt{-\det(g_{\mu\nu})}$ and  $L_{\rm min}$ is the Lagrangian of some minimally-coupled matter. The variation of the action with respect to $e^a_\mu$ yields the general equation of motion of $f(T)$ theories \cite{Maluf:1994ji,Ferraro:2012wp,Ferraro:2016wht}:
\be
-\ft{1}{2}e^\mu_a\, f +2\left(e^{-1}\partial_\sigma(ee^\lambda_a S_\lambda{}^{\sigma\mu})
+e^\lambda_a S_\rho{}^{\nu\mu}T^\rho{}_{\nu\lambda}\right) f_T
-2S_a{}^{\mu\lambda} \partial_\lambda T f_{TT}=8\pi\eta_{ab}\,e^b_\nu  T^{\mu\nu}_{\rm min}\,,
\ee
where $F_T=\frac{\partial f}{\partial T}$ and $f_{TT}=\frac{\partial^2 f}{\partial T^2}$ and $T^{\mu\nu}_{\rm min}$ is the matter energy momentum tensor. It can be demonstrated that Einstein gravity is a special case of $f(T)$ gravities, corresponding $f=-T -2\Lambda$.  Note that in this paper, we shall consider the cosmological constant as part of $f(T)$ rather than part of the matter energy-momentum tensor.

\section{Static ansatz and equations of motion}

\subsection{Spherically symmetric}
\label{vielansatz1}

The most general spherically-symmetric and static ansatz for the metric is given by
\be
ds^2 = -A(r) dt^2 + B(r) dr^2 + \rho(r)^2\big(d\theta^2 + \sin^2\theta\, d\phi^2\big)\,.
\label{met1}
\ee
In Einstein gravity, owing to the existence of local Lorentz symmetry of the tangent space, all the vielbein bases are equivalent and one typically chooses the most convenient diagonal basis for such a static and spherically-symmetric configuration.  In $f(T)$ theories, the choice matters. The diagonal vielbein basis, which we shall discuss in subsection \ref{tosy}, leads to no solutions for spherically symmetric ansatz in general $f(T)$.  One goal of this paper is to construct static and spherically-symmetric solutions that are asymptotic to the Minkowski spacetime.  We would like to further require that the vacuum of the $f(T)$ theory is the Minkowski spacetime in Cartesian coordinates where the vielbein is an identity matrix.
Specifically, the metric is
\be
ds^2 = -dt^2 + (dx^1)^2 + (dx^2)^2 + (dx^3)^2\,.
\ee
The diagonal vielbein $e^{\bar 0} = dt$ and $e^{\bar i}=dx^i$ is constant, i.e. $e^a_\mu=\delta^a_{\mu}$.  Therefore all the torsion components (\ref{torsiondef}) vanish identically, and hence the Minkowski spacetime with this vielbein choice is a solution of the $f(T)$ theory that satisfies $f(0)=0$.  This vacuum can be viewed as the most symmetric solution of the theory, we would like to construct spherically-symmetric solutions that are asymptotic to this vacuum.

Now using instead the spherically polar coordinates
\be
x^1=r\,\sin\theta\,\cos\phi\,,\qquad x^2=r\,\sin\theta\,\sin\phi\,,\qquad
x^3=r\,\cos\theta\,,
\ee
we have
\bea
e^{\bar 1} &=& \sin\theta\cos\phi\, dr - r \cos\theta\sin\phi\, d\theta - r \sin\theta \sin\phi\, d\phi\,,\nn\\
e^{\bar 2} &=& \sin\theta\sin\phi\, dr + r \cos\theta\sin\phi\, d\theta +
r\sin\theta\cos\phi\, d\phi\,,\nn\\
e^{\bar 3} &=& \cos\theta\, dr - r \sin\theta\, d\theta\,.\label{eisphere}
\eea
Since the $f(T)$ theory is invariant under general coordinate transformation,
the Minkowski spacetime in spherical polar coordinates must also be a solution of the $f(T)$ theory.  It is worth pointing out that if we make parity transformation such as $dr\rightarrow -dr$, we then have $e^{\rm i}\ne dx^i$ when written back in cartesian coordinates.  If follows that $e^{\bar i}$ in (\ref{eisphere}) with $dr\rightarrow -dr$, which has no effect on the metric, is no longer a solution in the $f(T)$ gravity.  We now promote the ansatz to replace $(r,dr)$ to a more general structure $(\rho(r), \sqrt{B(r)})$ and write the ansatz as
\be
e^a_\mu=\left(
  \begin{array}{cccc}
    \sqrt{A} & 0 & 0 & 0 \\
  0 & \sqrt{B}\,\sin\theta\cos\phi & \rho\,\cos\theta\cos\phi &
  -\rho\, \sin\theta\sin\phi \\
  0 & \sqrt{B}\,\sin\theta\sin\phi & \rho\,\cos\theta\sin\phi &
  \rho\,\sin\theta\cos\phi \\
  0 & \sqrt{B}\,\cos\theta & -\rho\sin\theta & 0 \\
  \end{array}
\right)\,,\label{vielbein}
\ee
where $e^{a}_\mu$ is the element in $(a+1)$'th row and $(\mu+1)$'s column, with $a=\bar 0,\bar 1,\bar 2,\bar 3$ and $\mu=0,1,2,3$.  This ansatz guarantees that the Minkowski spacetime, with $A=1=B$ and $\rho=r$, is the solution of the $f(T)$ theory. This ansatz is a special case of those involving three $SO(3)$ parameters \cite{Tamanini:2012hg}.

For the vielbein ansatz (\ref{vielbein}), we find that the torsion is given by
\be
T=-\fft{2(\rho'-\sqrt{B})}{B\rho}\Big( \fft{\rho'-\sqrt{B}}{\rho} + \fft{A'}{A} \Big)\,.
\ee
As we remarked earlier, when we take square roots of the metric functions to obtain the vielbein, we can have both positive and negative values, which are related to each other by some discrete improper Lorentz transformations.  The metric is invariant under such choices, but the equations of motion of $f(T)$ gravities can.  For example, in the above vielbein ansatz, we can perform a parity transformation by sending $\sqrt{B}\rightarrow -\sqrt{B}$. The metric is the same, but the torsion becomes
\be
T\rightarrow \tilde T=-\fft{2(\rho'+\sqrt{B})}{B\rho}\Big( \fft{\rho'+\sqrt{B}}{\rho} + \fft{A'}{A} \Big)\,.
\ee
The consequence is that for the Minkowski spacetime, corresponding to have $A=1=B$ and $\rho=r$, the vielbein choice (\ref{vielbein}) leads to vanishing torsion $T$, whilst the vielbein choice with $\sqrt{B}\rightarrow -\sqrt{B}$ leads to non-vanishing torsion $\tilde T$.  As we discussed earlier, the former yields the Minkowski spacetime as the vacuum, whilst latter has no Minkowski vacuum.  Solutions of the latter case were studied in \cite{Farrugia:2016xcw}.  In this paper, we shall only consider the former case and the static and spherically-symmetric solutions with the vielbein ansatz (\ref{vielbein}) that are asymptotic to the Minkowski spacetime will be constructed in section 4.

For the vielbein (\ref{vielbein}), we find that the equations of motion reduce to
\bea
&& F - F_T\,\Big( T + \fft{2}{B}\Big(
\frac{B' \rho'}{B \rho}+\frac{B}{\rho^2}-\frac{2 \rho''}{\rho}-\frac{\rho'^2}{\rho^2}
\Big)\Big) + f_{TT} T' \fft{4(\rho'-\sqrt{B})}{B\rho}=T^{\bar 0\bar 0}\,,\nn\\
&&-F - \fft{F_T}{B}\Big( \frac{2 A''}{A}-\frac{A' B'}{A B}-\frac{2 \sqrt{B} A'}{A \rho}+\frac{4 A' \rho'}{A \rho}-\frac{A'^2}{A^2}+\frac{4 B}{\rho^2}-\frac{4 \sqrt{B} \rho'}{\rho^2}\Big)\nn\\
&&\qquad\qquad - f_{TT} T' \fft{2A'}{AB}  =
T^{\bar0\bar0} + 2 T^{\bar2\bar2} = T^{\bar0\bar0} + 2 T^{\bar 3\bar3}\,,\nn\\
&&-F + \fft{2f_{T}}{B}\Big(\frac{\sqrt{B} A'}{A \rho}-\frac{2 A' \rho'}{A \rho}+\frac{2 \sqrt{B} \rho'}{\rho^2}-\frac{2 \rho'^2}{\rho^2}
\Big)=T^{\bar 1\bar 1}\,.\label{eom1}
\eea
It is worth checking that the above equations of motion are consistent with $\nabla_\mu T^{\mu\nu}_{\rm min}=0$.  In particular, for vacuum solutions with $T^{\mu\nu}_{\rm min}=0$, the quantity $H=T^{\bar 1\bar 1}$ above can be viewed as the first integral of the remaining two second-order differential equations, satisfying
\be
H' = - \Big( \fft{A'}{2A} + \fft{2\rho'}{\rho}\Big) H\,.
\ee
It is thus consistent to set $H=0$.

\subsection{Toroidally symmetric}
\label{tosy}

The most general static ansatz with spherical, toroidal or hyperbolic symmetries takes the form
\be
ds^2 = -A(r) dt^2 + B(r) dr^2 + \rho(r)^2\Big(\fft{dx^2}{1-k x^2} + (1-k x^2) dy^2\Big)\,,
\ee
with $k=1,0,-1$ respectively.  In this subsection, we consider the diagonal vielbein basis
\be
e^{\bar 0} = \sqrt{A}\, dt\,,\qquad e^{\bar 1}=\sqrt{B}\, dr\,,\qquad
e^{\bar 2} = \fft{\rho}{\sqrt{1-k x^2}}\, dx\,,\qquad
e^{\bar 3} = \rho \sqrt{1-k x^2}\,dy\,.
\ee
For this choice of the vielbein, the torsion is given by
\be
T=-\fft{1}{B}\Big(\fft{A' \rho'}{A\rho} + \fft{\rho'^2}{2\rho^2}\Big)\,.
\ee
Unlike the previous example, the equations of motion are independent of the sign choice of the vielbein.  It is well known that for vacuum solution or diagonal matter energy momentum tensor, the equation of motion in the $(\bar 1,\bar 2)$ direction implies that
\be
k f_{TT}=0\,.
\ee
It is satisfied either with $f_{TT}=0$, giving rise to Einstein gravity, or with $k=0$, corresponding to toroidal topology with the metric becomes
\be
ds^2 = -A(r) dt^2 + B(r) dr^2 + \rho(r)^2(dx^2 + dy^2)\,.
\ee
In this paper, we consider the latter case, and the equations of motion reduce to
\bea
&& F - F_T\,\Big( T + \fft{2}{B}\Big(
\frac{B' \rho'}{B \rho}-\frac{2 \rho''}{\rho}-\frac{\rho'^2}{\rho^2}
\Big)\Big) + 4f_{TT} \fft{T'\rho'}{B\rho}=T^{\bar 0\bar 0}\,,\nn\\
&&-F - \fft{F_T}{B}\Big( \frac{2 A''}{A}-\frac{A' B'}{A B}+\frac{4 A' \rho'}{A \rho}-\frac{A'^2}{A^2}\Big)- f_{TT} \fft{2A' T'}{AB}  =
T^{\bar0\bar0} + 2 T^{\bar2\bar2} = T^{\bar0\bar0} + 2 T^{\bar 3\bar3}\,,\nn\\
&&-F -\fft{4f_{T}}{B}\Big(\frac{A' \rho'}{A \rho}+\frac{\rho'^2}{\rho^2}
\Big)=T^{\bar 1\bar 1}\,.
\eea
In section \ref{adslifs}. we construct AdS planar black holes in $f(T)$ gravities.

\subsection{Energy-momentum tensor in Einstein gravity}

It is of interest to compare Einstein gravity and $f(T)$ gravity. In particular we would like to know whether solutions that violate the null energy conditions in Einstein gravity can arise as good solutions in $f(T)$ gravity.  Since in this paper, we consider static diagonal metrics, the corresponding energy momentum tensor in Einstein gravity is also diagonal, namely
\be
T^{ab}_{\rm Ein}= {\rm diag}\{\rho_{\rm E},\, p^1_{\rm E},\, p^2_{\rm E},\, p^3_{\rm E}\}\,,
\ee
where
\bea
\rho_{\rm E} &=& \fft{1}{B}\Big(\frac{B' \rho'}{B \rho}+\frac{k\,B}{\rho^2}-\frac{2 \rho''}{\rho}-\frac{\rho'^2}{\rho^2}\Big)\,,\nn\\
p^1_{\rm E} &=& \fft{1}{B}\Big(\frac{A' \rho '}{A \rho }-\frac{k\,B}{\rho ^2}+\frac{\rho '^2}{\rho ^2}\Big)\,,\nn\\
p^2_{\rm E} &=& p^3_{\rm E} = \fft{1}{B}\Big(\frac{A''}{2 A}-\frac{A' B'}{4 A B}+\frac{A' \rho '}{2 A \rho }-\frac{A'^2}{4 A^2}-\frac{B' \rho '}{2 B \rho }+\frac{\rho ''}{\rho }
\Big)\,.\label{rhop}
\eea
These quantities allow us to check the energy conditions for solutions of $f(T)$ gravities had they been constructed in Einstein gravity.

\section{Asymptotically Minkowski solutions}

In this section, we consider metric and vielbein ansatz (\ref{met1}) and (\ref{vielbein}), and solve the equations of motion (\ref{eom1}) for vanishing energy-momentum tensor, $T^{\mu\nu}_{\rm min}=0$.

\subsection{Minkowski vacuum in spherical polar coordinates}

It follows from (\ref{met1}) that the Minkowski vacuum, with the Euclidean $\mathbb R^3$ written in spherical polar coordinates,
is given by
\be
A=1=B\,,\qquad \hbox{and}\qquad \rho=r\,.\label{minsol}
\ee
It is clear that the above solves the equations in (\ref{eom1}) with $T^{\mu\nu}_{\rm min}=0$
provided that the function $F$ must satisfy
\be
f(T)=f(0)=0\,.
\ee
This is because the torsion $T=0$ the Minkowski vacuum (\ref{minsol}).  Thus $f(T)$ theories of the type (\ref{FT1}) considered in the introduction all admit such a vacuum. One goal of this paper is to study how the $\alpha$ term affects the asymptotic falloffs from this vacuum.

\subsection{No global AdS vacuum}
\label{ngav}

The metric of AdS vacuum in global coordinates can be written as (\ref{met1}) with
\be
A=g^2 r^2 + 1=\fft{1}{B}\,,\qquad\rho=r\,.
\ee
Substituting this into (\ref{eom1}) and we find that equations can only be satisfied provided that $f_{TT}=0$, which reduces to Einstein gravity.  However, there can exist nevertheless asymptotically {\it locally} AdS solutions.  It should be pointed that AdS spacetimes in global coordinates can be constructed in $f(T)$ gravities with appropriate vielbein ansatz.
(See, e.g.~\cite{Tamanini:2012hg}.)

\subsection{Special exact solutions}
\label{ses}

The equations (\ref{eom1}) can be solved up to a quadrature, provided that $f(T)$ has a double root, namely
\be
F=(T-\lambda)^2 U(T)\,.\label{quadraticF}
\ee
It can be seen easily that in this case all the equations can be solved with
\be
T=\lambda\,.
\ee
This implies that
\be
-\fft{2\sqrt{f}\,(\sqrt{f}-1)\,h'}{r h} - \fft{2(\sqrt{f}-1)^2}{r^2}=\lambda\,.\label{hfeom1}
\ee
Here, we have made a coordinate gauge $\rho=r$ and define $(A,B)=(h,f^{-1})$, so that the metric is expressed as
\be
ds^2 = - h(r)\, dt^2 + \fft{dr^2}{f(r)} + r^2 (d\theta^2 + \sin^2\theta\, d\phi^2)\,.
\label{hfmet1}
\ee
(There should be no confusion between the metric function $f(r)=1/g_{rr}$ with the $f(T)$ function.) The equation (\ref{hfeom1}) implies that
\be
h=\exp\Big(-\int dr\, \fft{2f-4\sqrt{f} + \lambda r^2 +2}{2r(f-\sqrt{f})}\Big)\,.
\ee
In this subsection, we focus our attention on solutions that are asymptotically to the Minkowski spacetime.  Thus we set $\lambda=0$, and we have
\be
f=\fft{h^2}{(h + r h')^2}\,.\label{hfeom2}
\ee
We find following classes of exaction solutions.

\subsubsection{Extremal black holes}

Asymptotically-flat black holes can emerge, but they are all extremal.  It follows from (\ref{hfeom2}) that when $r=r_0$ is a single root of $h$, it is a double root for $f$. We present two concrete examples:
\bea
\hbox{Solution 1}:&&\qquad h=f=\Big(1 - \sqrt{\fft{r_0}{r}}\Big)^2\,;\nn\\
\hbox{Solution 2}:&&\qquad h=1 - \fft{2M}{r}\,,\qquad f=\Big(1 - \fft{2M}{r}\Big)^2\,.\label{bh12}
\eea
Both solutions describe extremal black holes with zero temperature.  The first solution has a slower falloff than the Schwarzschild black hole, whilst the second has the same $1/r$ falloff from the asymptotic Minkowski spacetime.

      It is of interest to study the energy-momentum tensor of the solutions
(\ref{bh12}) if they were embedded in Einstein gravity.  It follows from (\ref{rhop}) that we have
\bea
\hbox{Soluton 1}: &&\qquad T_{\rm Ein}^{ab}=\sqrt{\fft{r_0}{r^5}}\,{\rm diag}\{1,-1,\ft14,\ft14\}\,;\nn\\
\hbox{Solution 2}: &&\qquad
T_{\rm Ein}^{ab}=\fft{M}{r^4}{\rm diag}\left\{4M,-2r,r-M,r-M)\right\}\,.
\eea
Thus we see that if they were to be embedded in Einstein gravity, the first solution satisfies the null energy condition whilst the second does not.

\subsubsection{Dark wormholes}
\label{darkworm}

Smooth solutions can also arise from (\ref{hfeom2}), when $f$ has a single root and $h$ has a branch cut.  As a concrete example, we present a solution
\be
f=1-\fft{r_0^2}{r^2}\,,\qquad h=\ft12 (1 + \sqrt{f})\,.\label{wh1}
\ee
It is not the most convenient coordinate system to study the global structure. Making a coordinate transformation $r^2= \eta^2 + r_0^2$, we have
\be
ds^2=- A\, dt^2 + d\eta^2 + (\eta^2 + r_0^2) (d\theta^2 +\sin^2\theta\, d\phi^2)\,,\qquad
A=\ft12 \Big(1 + \fft{\eta}{\sqrt{\eta^2 + r_0^2}}\Big)\,.\label{smooth1}
\ee
In this paper, we refer $\eta$ as the ``proper'' radial coordinate.  The low-lying curvature polynomial invariants are given by
\bea
R &=& - \fft{3r_0^2}{2(\eta^2 + r_0^2)^2}\,,\nn\\
R^{\mu\nu}R_{\mu\nu} &=& -\fft{\eta(6\eta^2 + 7r_0^2)}{2(\eta^2 + r_0^2)^{\fft72}} +
\fft{24\eta^4 + 40 \eta^2 r_0^2 + 25 r_0^4}{8 (\eta^2 + r_0^2)^4}\,,\cr
R^{\mu\nu\rho\sigma}R_{\mu\nu\rho\sigma} &=& -\fft{\eta(6\eta^2 -r_0^2)}{2(\eta^2 + r_0^2)^{\fft72}} +\fft{48\eta^4 + 16 \eta^2 r_0^2 + 49 r_0^4}{4 (\eta^2 + r_0^2)^4}\,.
\eea
Thus the curvature has local power-law singularity at $\eta ={\rm i} r_0$.  The curvatures are all regular for real $\eta$.  In particular, all the curvature polynomials vanish at the two asymmetric asymptotic regions of $\eta\rightarrow \pm \infty$.  Thus, for the above solution, proper radial coordinate $\eta$ runs smoothly from $+\infty$ to $-\infty$, with
\bea
\eta\rightarrow +\infty:&& A= 1 -\fft{r_0^2}{4\eta^2} + {\cal O}(\eta^{-4})\,,\nn\\
\eta\rightarrow -\infty:&& A= \fft{r_0^2}{4\eta^2} + {\cal O}(\eta^{-4})\,.
\eea
The solution is asymptotically flat with vanishing mass term as $\eta\rightarrow +\infty$. The behavior at $\eta\rightarrow -\infty$ is intriguing. The curvature vanishes as $r\rightarrow -\infty$, indicating it is locally flat.  Furthermore, we have $A\sim 1/\eta^2$ for large negative $\eta$, which indicates there exists a horizon located at $\eta\rightarrow -\infty$.  The spacetime is geodesically complete for $\eta\in (-\infty,\infty)$ and has no curvature singularity, and the time coordinate $t$ is globally defined.
Such a smooth spacetime configuration is very different from solitons where the proper radius lies in the semi-infinity region $\eta\in [0,\infty)$. The solution resembles instead a wormhole, connecting two asymptotic regions, with the wormhole radius $r_0$.  In fact, for any constant slice of time, the spatial section is precisely the same as  the Ellis wormhole \cite{Ellis:1973yv}, namely
\be
ds_3^2 = d\eta^2 + (\eta^2 + r_0^2)\, (d\theta^2 +\sin^2\theta\, d\phi^2)\,.
\ee
However, there is a key difference.  In a typical wormhole, there exists a turning point $\eta=\eta_0$ such that $A'(\eta_0)=0$, whilst there is no such a turning point in our solution.  In fact $-g_{tt}$ in our solution is a monotonically increasing function as $\eta$ runs from $-\infty$ to $+\infty$.  In particular, signals sent from $-\infty$ through the wormhole is infinitely red-shifted.  The wormhole observed from the $\eta>0$ region is somewhat dimmed by the red-shift and we call such a spacetime configuration as a dark wormhole.  Since the total ``black'' region is at $r=-\infty$, the wormhole is darker than the Ellis wormhole, but not as black like a black hole.

It is worth pointing out that such a solution cannot arise in Einstein gravity since it violates the null energy condition, as in the case of the Ellis wormhole.  In fact, it follows from (\ref{rhop}) that we have
\be
\rho_{\rm E} + p_{\rm E}^1 = -\fft{1}{r^2}(1-\sqrt{f})(2 + \sqrt{f})\,,
\ee
which is negative, since $|\sqrt{f}|<1$.

\begin{figure}[htp]
\begin{center}
\includegraphics[width=200pt]{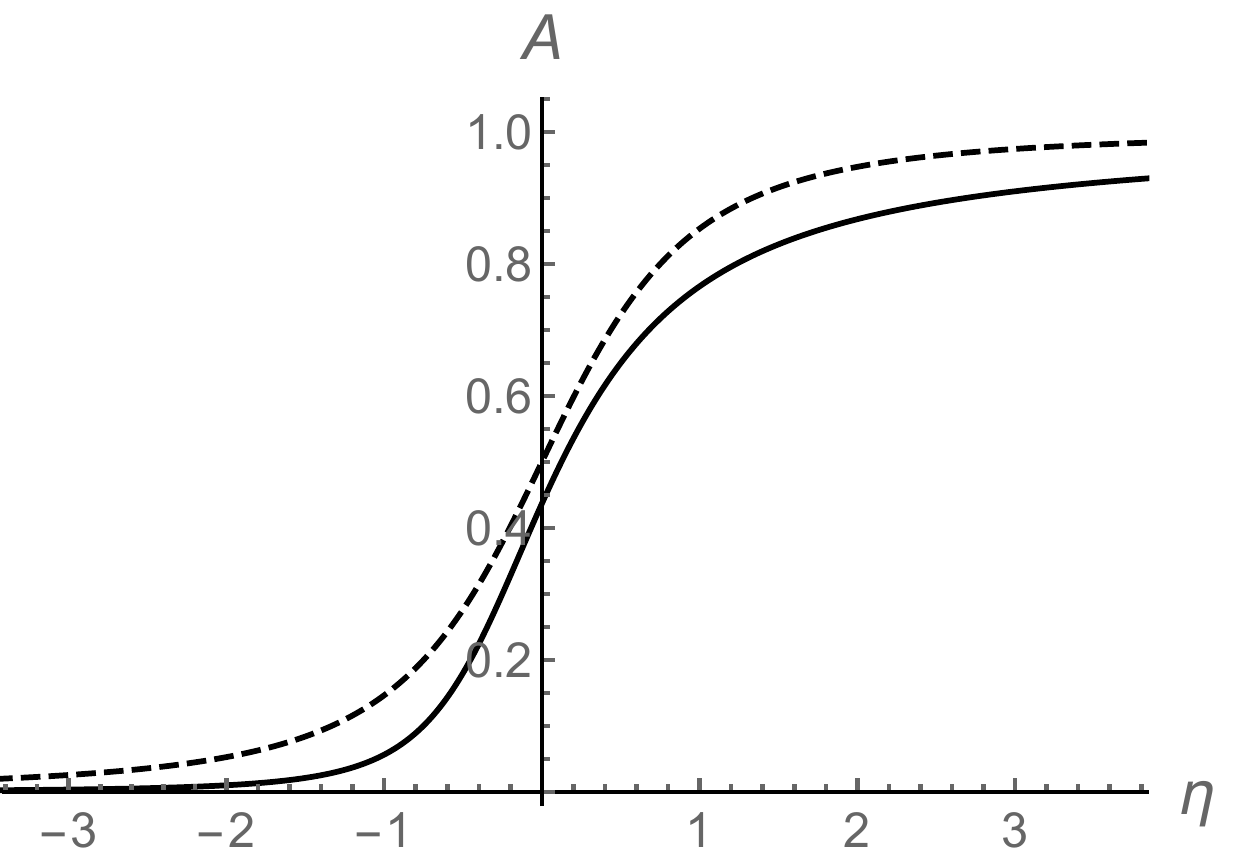}
\end{center}
\caption{\setstretch{1.0}\small\it Plots of $A=-g_{tt}$ as a function of proper radial coordinate $\eta$. The dashed line is for the solution (\ref{smooth1}) with $r_0=1$. The solid line is for the solution (\ref{smooth2}) with $M=1/4$.  For both cases, the function $A$ runs from 0 to 1 as $\eta$ runs from $-\infty$ to $+\infty$.} \label{figure1}
\end{figure}

We now present another dark wormhole solution, but with a slower $1/r$-falloff mass term at large positive $r$. It is given by
\be
 f=1- \fft{2M}{r}\,,\qquad h=\ft14 (\sqrt{f} +1)^2\,.\label{wh2}
\ee
This solution would also violate the null energy condition in Einstein gravity, with
\be
\rho_{E} + p_{\rm E}^1 = - \fft{2M}{r^3\big(1 + \sqrt{f}\big)}\,.
\ee
The relation between the proper radius $\eta$ and $A$ is given by
\be
\eta=\ft12M\Big(\fft{1}{1-\sqrt{A}}-\fft{1}{\sqrt{A}}\Big) -
M\log\fft{1}{M}\big(\fft{1}{\sqrt{A}}-1\big)\,.\label{smooth2}
\ee
Thus we see that as $\eta$ runs from $-\infty$ to $+\infty$, the function $A$ runs from 0 to 1.  In Fig.~\ref{figure1}, we give plots of $A=-g_{tt}$ as a function of the proper radius $\eta$, which indicates that the spacetime is smooth and geodesically complete.

The dark wormhole behavior can also be understood directly using the original radial coordinate $r$.  For solutions (\ref{wh1}) and (\ref{wh2}), as the metrics run from Minkowski spacetime at $r=\infty$ to the wormhole throat $r=r_0$ where $f(r_0)=0$, the continuity for the metric functions and their derivatives imply that the $\sqrt{f}$ term in $h$ reverses sign and becomes $-\sqrt{f}$ at $r=r_0$, leading to the asymmetric asymptotic region where $g_{tt} \rightarrow 0$.

Such a dark wormhole solution was in fact first constructed in non-minimally coupled Einstein vector theory.  It was shown in \cite{Geng:2015kvs} that the four-dimensional Lagrangian
\be
{\cal L}=\sqrt{-g} \Big(R - 2\Lambda_0 - \ft14 F^2 +\gamma (G_{\mu\nu} + \Lambda_0 g_{\mu\nu}) A^\mu A^\nu\Big)\,,
\ee
admits the solution (\ref{hfmet1}) with
\be
f=-\ft13\Lambda_0 r^2 + 1 - \fft{2M}{r}\,,\qquad
h=\alpha + \ft12 \Big(f + \sqrt{f(4\alpha + f)}\Big)\,.
\ee
Analogous to solutions in the $f(T)$ theories we constructed, the above solution of Einstein-vector theory also describes a dark wormhole that connects two asymptotic AdS ($\Lambda_0<0$) or Minkowski ($\Lambda_0=0$) regions.  It is tantalizing to notice that such solutions can emerge in so different theories.

\subsection{General features of the solutions}

As has been shown previously, when the $f(T)$ satisfies $f(0)=0$ and that $f_T(0)$ and $f_{TT}(0)$ are regular, then the theory admits Minkowski vacuum.  In this subsection, we study the general features of the solutions for the class of theories (\ref{FT1}) mentioned in the introduction. Furthermore, we shall focus on the $n=2$ case.

\subsubsection{Small-$\alpha$ solutions}

First we consider the case with small $\alpha$. This is a reasonable assumption under the philosophy that $f(T)$ theories are introduced to provide small modifications to Einstein gravity. When $\alpha=0$, the spherically-symmetric and static solution is the Schwarzschild black hole.  We adopt the coordinate gauge that $r$ is the radius of the $S^2$, and the metric takes the form (\ref{hfmet1}).  We consider perturbation away from the Schwarzschild black hole solution with $\alpha$ as the perturbation parameter.  In other words, we consider
\be
h=\bar f\,  (1 + \alpha \tilde h) + {\cal O}(\alpha^2)\,,\qquad
f=\bar f\,  (1 + \alpha \tilde f) + {\cal O}(\alpha^2)\,,\qquad \bar f=1-\fft{2M}{r}\,.
\ee
For (\ref{FT1}) with $n=2$, we find
\bea
\tilde h &=& \fft{32\sqrt{\bar f}}{3M^2} + \frac{6 M^2+39 M r-64 r^2}{6 M^2 r^2} - \fft{c}{r \bar f} + \fft{(1-\bar f)\log \bar f}{2M^2\bar f}\,,
\nn\\
\tilde f &=& \fft{21r-50M}{2Mr^2} -\frac{16 \left(3 M^2-7 M r+2 r^2\right)}{3 Mr^3\sqrt{\bar f}} - \fft{c M + 2}{M r\bar f} +\frac{\log (\bar f )}{M r\bar f}\,,
\eea
where $c$ is an integration constant.  It is clear that the perturbation diverges on the horizon of the background black hole and the integration constant $c$ cannot be used to removed all the divergence.  Thus there is no new black hole in the vicinity of the Schwarzschild black hole of Einstein gravity.  The asymptotic falloffs due to the $\alpha$ contribution is $1/r^5$, much faster than the $1/r$ falloff associated with the Schwarzschild black hole. This implies that in the weak field region, the $\alpha$-term contribution is negligible.

\subsubsection{Large-$\alpha$ solutions}

When $\alpha$ becomes infinite, the theory (\ref{FT1}) reduces to those discussed in the subsection \ref{ses}, where some special exact solutions were constructed.  We can perform $1/\alpha$ expansion for large $\alpha$ and find that the solutions are given by
\bea
h = \bar h \Big( 1 + \fft{1}{\alpha} \tilde h_1 + \fft{1}{\alpha^{3/2}} \tilde h_2
+ \cdots\Big)\,,\quad
f = \fft{\bar h^2}{(\bar h+ r\bar h')^2}\Big(1 + \fft{1}{\alpha} \tilde f_1 + \fft{1}{\alpha^{3/2}} \tilde f_2 + \cdots
\Big)\,,
\eea
with
\be
\tilde f_1 = \frac{r \left(\bar h + r\bar h'\right)}{2 \bar h'}-\frac{2 r \bar h }{\bar h + r \bar h'}\,\tilde h_1'\,,\qquad
\tilde f_2 = -\fft{2r \bar h}{\bar h + r\bar h'}\, \tilde h_2'\,.
\ee
The solutions have analogous properties discussed in subsection \ref{ses}.
\subsubsection{Generic $\alpha$}

For the general case of (\ref{FT1}), we adopt the coordinate gauge that $r$ is the radius of the $S^2$, and the metric takes the form (\ref{hfmet1}).  For large $r$, we find
\be
h=1 - \fft{2M}{r} - \fft{n 2^n M^{2n-1}}{(4n-3)}\fft{\alpha}{r^{4n-3}} + \cdots\,,\qquad
f=1 - \fft{2M}{r} - n 2^n M^{2n-1}\fft{\alpha}{r^{4n-3}} + \cdots\,.
\ee
The fast falloffs indicates that in the long wavelength infra-red region, the $f(T)$ gravities deviate from Einstein gravity slightly.  This implies that $f(T)$ gravities can pass the solar-system tests easily, without needing to require that the parameter $\alpha$ be particularly small.

The structure in the middle however is rather unclear.  It can be shown that the standard event horizon, with the near-horizon expansion
\be
h=h_1 (r-r_0) + h_2 (r-r_0)^2 + \cdots\,,\qquad
f=f_1 (r-r_0) + f_2 (r-r_0)^2 + \cdots\,,
\ee
does not satisfy the equations of motion.  Neither the extremal case can arise. We appeal to the numerical analysis and integrate the equations from asymptotic infinity to the middle.

\begin{figure}[htp]
\begin{center}
\includegraphics[width=250pt]{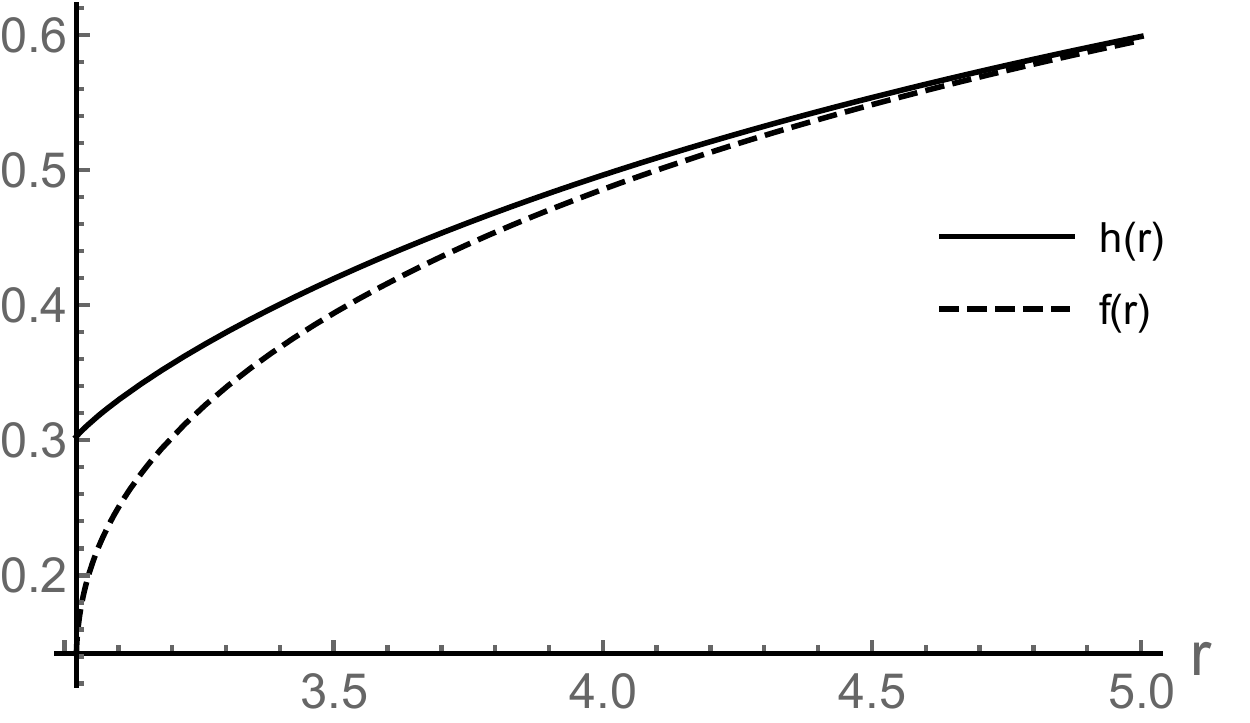}
\end{center}
\caption{\setstretch{1.0}\small\it Plots  of $h$ and $f$ for $n=2$, $M=1$ and $\alpha =1$.
$(h,f)$ approaches 1 at $r\rightarrow \infty$, and $(0.3028, 0.1422)$ at $r\sim 3.0218$, with
$(h',f')$ and $(h'',f'')$ diverging.  This suggests there is a branch-cut singularity at $r\sim 3.0218$.
\label{figure2}}
\end{figure}

Concretely, we consider $n=2$, and expand the solutions to sufficiently higher orders to increase accuracy, namely
\bea
h &=& 1-\frac{2 M}{r}-\frac{8 \alpha M^3}{5 r^5}-\frac{58\alpha M^4}{15 r^6}-\frac{54\alpha M^5}{7 r^7}-\frac{409\alpha M^6}{28 r^8} + \cdots\,,\nn\\
f &=& 1-\frac{2 M}{r}-\frac{8 \alpha M^3}{r^5} -\frac{62 \alpha M^4}{5 r^6}-\frac{58 \alpha M^5}{3 r^7}-\frac{219 \alpha M^6}{7 r^8}  + \cdots\,.
\eea
Choosing parameter $M=1=\alpha$, and integrate from $r=1000$ to zero, we found that the equations become singular at $r\sim 3.0218$, with $(h,f)\rightarrow (0.3028, 0.1422)$, but $(h',f')$ and $(h'',f'')$ become diverge, indicating that there is a branch-cut singularity at $r\sim 3.0218$.  We present the plots of $(h,f)$ in Fig.~\ref{figure2}.  We examine various points in the parameter space and find that this is the general pattern.  It is worth noting that a branch-cut singularity of the metric function can sometime indicate the existence of dark wormhole discussed earlier, but the wormhole configuration in metric ansatz (\ref{hfmet1}) requires $f\rightarrow 0$ at certain wormhole throat $r\rightarrow r_0$.

\section{Asymptotically locally AdS solitons and black holes}

In subsection \ref{ngav}, we showed that within our vielbein ansatz, the AdS vacuum in global coordinates was not a solution of $f(T)$ gravities.  In this section, we demonstrate that there exist asymptotically locally AdS solitons.  The theory we consider in this section is (\ref{FT1}) augmented with a cosmological constant, namely
\be
f(T) = - T + \alpha T^n - 2\Lambda_0\,,\label{FT2}
\ee
where the cosmological constant is negative, parameterized as
\be
\Lambda_0=-3 g^2\,.
\ee

\subsection{Solitons with small $\alpha$ solutions}

When $\alpha=0$, the $f(T)$ theory reduces to Einstein gravity and the AdS spacetime in global coordinates is the vacuum, given by (\ref{hfmet1}) with $h=f=f_0\equiv g^2 r^2 + 1$.  In this subsection, we consider small $\alpha$, the corresponding solution in the $f(T)$ can be viewed as a perturbation away from the AdS with the $\alpha$ as its expansion parameter.  We thus consider the metric (\ref{hfmet1}) with the ansatz
\be
h=f_0\, (1 + \alpha \tilde h + {\cal O}(\alpha^2))\,,\qquad
f=f_0\, (1 + \alpha \tilde f + {\cal O}(\alpha^2))\,.
\ee
We find that up to and including the linear order of $\alpha$, the $(\tilde h,\tilde f)$ functions are given by two quadratures
\be
\tilde f = \fft{U}{r f_0}\,,\qquad \tilde h = \tilde f + V\,,
\ee
where
\bea
U' &=& (-2g^2)^{n-1} \left(\sqrt{f_0}-1\right)^n \left(\sqrt{f_0}+1\right)^{1-n} \left(3 \sqrt{f_0}+2\right)^{n-2} f_0^{-\frac{n}{2}} \Big(9(2 n-1)f_0^{3/2}\nn\\
&&+(2n+3)(4n-1)f_0 +2\left(4 n^2-5 n+4\right) \sqrt{f_0} +4 (n-1)^2\Big)\,,\nn\\
V' &=& -\fft{(-2)^{n+1} (n-1) n \left(2f_0+2 \sqrt{f_0}+1\right)}{ g^{1-2 n} f_0^{\frac{n}{2}+1}} \left(\frac{\sqrt{f_0}-1}{\sqrt{f_0}+1}\right)^{n-\frac{1}{2}} \left(3 \sqrt{f_0}+2\right)^{n-2}\,.
\eea
These quadratures can be integrated out for integer $n$. For example, when $n=2$, we have
\bea
U&=&\frac{16 \left(2 g^2 r^2+3\right) \sqrt{g^2 r^2+1}}{r}-\frac{2 \left(9 g^4 r^4+8 g^2 r^2+24\right)}{r}\nn\\
&&-8 g {\rm arctan}(g r)-32 g {\rm arcsinh}(g r)\,,\nn\\
V &=& \frac{8 \left(3 g^2 r^2+2\right) \left(g^2 r^2 \left(\sqrt{g^2 r^2+1}-2\right)+2 \left(\sqrt{g^2 r^2+1}-1\right)\right)}{r^2 \left(g^2 r^2+1\right)^{3/2}} \nn\\
&&+ 8 g^2 \log\Big(\fft{(\sqrt{g^2r^2 + 1}+1)^2}{4(g^2r^2+1)}\Big)\,.
\eea
For $n=3$, we have
\bea
U &=& 180 g^6 r^3+672 g^4 r+48 g^3 {\rm arctan}(g r)+832 g^3 {\rm arcsinh}(g r)+\frac{2016 g^2}{r}\nn\\
&&-\frac{32 \left(54 g^6 r^6+235 g^4 r^4+209 g^2 r^2+40\right)}{3 r^3 \sqrt{g^2 r^2+1}}+\frac{1280}{3 r^3}\,,\nn\\
V &=& \frac{32 \left(27 g^6 r^6+65 g^4 r^4+48 g^2 r^2+12\right)}{r^4 \left(g^2 r^2+1\right)^{3/2}}-\frac{16 \left(22 g^6 r^6+91 g^4 r^4+84 g^2 r^2+24\right)}{r^4 \left(g^2 r^2+1\right)}\nn\\
&&- 144 g^4 \log\Big(\fft{(\sqrt{g^2r^2 + 1}+1)^2}{4(g^2r^2+1)}\Big)\,.
\eea
For $n=4$, we have
\bea
U &=& 384 g^5 {\rm arctan}(g r)-13056 g^5 {\rm arcsinh} (g r)\cr
&&+\frac{128 \left(270 g^8 r^8+1883 g^6 r^6+2389 g^4 r^4+1004 g^2 r^2+168\right)}{5 r^5 \sqrt{g^2 r^2+1}}\cr
&&-\frac{8 \left(945 g^{10} r^{10}+8865 g^8 r^8+39480 g^6 r^6+45920 g^4 r^4+17408 g^2 r^2+2688\right)}{5 r^5 \left(g^2 r^2+1\right)}\,,\cr
V &=& \frac{64 \left(42 g^{10} r^{10}+351 g^8 r^8+843 g^6 r^6+856 g^4 r^4+392 g^2 r^2+64\right)}{r^6 \left(g^2 r^2+1\right)^2}\cr
&&-\frac{128 \left(69 g^8 r^8+271 g^6 r^6+342 g^4 r^4+180 g^2 r^2+32\right)}{r^6 \left(g^2 r^2+1\right)^{3/2}}\cr
&& + 960 g^6 \log\Big(\fft{(\sqrt{g^2r^2 + 1}+1)^2}{4(g^2r^2+1)}\Big)\,.
\eea
In all these examples, we have chosen the integration constants such that the functions $(\tilde h, \tilde f)$ are regular in the whole $r\in [0,\infty)$ region. The effective cosmological constant at large $r$ is
\be
\Lambda_{\rm eff}=-3g^2 \Big(1 + (2n-1)(-6)^{n-1} g^{2n-2}\alpha\Big)\,.
\ee
On the other hand, as $r\rightarrow 0$, the metric becomes like the AdS solution with $\Lambda_0=-3g^2$.  Thus the solution describes smooth soliton that flows from the AdS vacuum of $\Lambda=\Lambda_0$ in the middle to the asymptotically locally AdS spacetimes of $\Lambda=\Lambda_{\rm eff}$.

\subsection{Solitons for generic $\alpha$}

In the previous subsection, we showed that smooth soliton solutions existed for small $\alpha$.  In this section, we employ numerical analysis to show that such solutions exist for generic $\alpha$.  In order to set up the boundary condition for numerical analysis, we assume that the metric is regular at the middle $r=0$, a requirement for a smooth soliton.  For the small-$r$ geometry, we assume that the metric functions have regular Taylor expansion. This ensures that the metric is regular and geodesic complete in the middle.  Substituting the ansatz into the equations of motion, we find
\bea
h &=& h_0 \Big(1 + g^2 r^2 - \ft52 \alpha g^6 r^4 + \ft{1}{168} \alpha g^8 (
271 + 4200 \alpha g^2) r^6 + \cdots\Big)\,,\nn\\
f &=& 1 + g^2 r^2 - 5 \alpha g^6 r^4 + \ft1{56} \alpha g^8 (
151+4200\alpha g^2) r^6 + \cdots\,.\label{solitontaylor}
\eea
Here $h_0$ is an integration constant that can be absorbed into the time scaling. Note that there is no non-trivial free parameters in the expansion.  When $\alpha=0$, the theory reduces to Einstein gravity and the solution is the AdS in global coordinates.  Thus we see that in the $r=0$ middle, the solution behaves like the middle of the AdS. The lacking of an extra free parameter indicates that such configuration cannot arise in Einstein gravity. Indeed, it follows from (\ref{rhop}) that we have
\be
\rho_{\rm E} + p_{\rm E}^1 = 10\alpha g^6 r^2 + {\cal O}(r^4)\,,\qquad
\rho_{\rm E} + p_{\rm E}^1 = -5 \alpha g^6 r^2 + {\cal O}(r^4)\,.
\ee
Thus the null-energy condition cannot be satisfied if one were to embed such a solution in Einstein gravity.

In the asymptotic region, we find that the metric functions become
\bea
h &=& \tilde g^2 r^2 + \fft{1 + 52\alpha \tilde g^2}{1 + 36\alpha \tilde g^2} +
\fft{32\alpha \tilde g (1 + 56\alpha \tilde g^2 + 1104 \alpha^2 \tilde g^4)}{(1+
36\alpha \tilde g^2)^3}\fft{\log r}{r} - \fft{2M}{r} + \cdots\,,\cr
f &=& \tilde g^2 r^2 + \frac{32 \alpha  \tilde g^3}{36 \alpha  \tilde g^2+1} r +
\frac{62784 \alpha ^3 \tilde g^6+4272 \alpha ^2 \tilde g^4+92 \alpha  \tilde g^2+1}{\left(36 \alpha  \tilde g^2+1\right)^3}\nn\\
&& -\frac{32 \alpha  \tilde g \left(1104 \alpha ^2 \tilde g^4+56 \alpha  \tilde g^2+1\right) }{\left(36 \alpha  \tilde g^2+1\right)^3} \fft{\log r}{r} - \fft{2M}{3 \left(36 \alpha  \tilde g^2+1\right)^5}\,{\fft1 r}\nn\\
&&+\frac{128 \alpha  \tilde g \left(1368576 \alpha ^4 \tilde g^8+141696 \alpha ^3 \tilde g^6+5424 \alpha ^2 \tilde g^4+96 \alpha  \tilde g^2+1\right)}{3 \left(36 \alpha  \tilde g^2+1\right)^5}\,\fft{1}{r} + \cdots\,,\label{solitonasym}
\eea
where the effective cosmological constant $\Lambda_{\rm eff}=-3\tilde g^2$ is related to $\Lambda_0$ by
\be
\Lambda_0 = -3\tilde g^2 (1 + 18\alpha \tilde g^2)=\Lambda_{\rm eff} (1 - 6\alpha \Lambda_{\rm eff})\,.
\ee
Thus for given $\Lambda_0$, we have
\be
\Lambda_{\rm eff}^\pm = \fft{1 \pm \sqrt{1-24\alpha \Lambda_0}}{12\alpha}\,.\label{lameffpm}
\ee
In this paper, we consider negative cosmological constants with $\Lambda_0=-3g^2$ and $\Lambda_{\rm eff}=-3\tilde g^2$, then we have
\be
\tilde g^2_{\pm} = -\fft{1\pm \sqrt{1 + 72\alpha g^2}}{36\alpha}\,.
\ee
For positive $\alpha$, only $\tilde g_-$ is real and it can be shown that $\tilde g_-^2 <g^2$. For $-g^2/(72)<\alpha <0$, we have $\tilde g_{+}^2> \tilde g_-^2>g^2>0$.

The question now remains whether the regular metric (\ref{solitontaylor}) in the middle can smoothly connect to the asymptotic region (\ref{solitonasym}).  We employ numerical method to establish this.  We use the Taylor expansion (\ref{solitontaylor}) at some small $r$, (e.g.~$r=0.1$), and integrate out to large $r$, (e.g.~$r=10^4$.)  Our numerical results indicate that for $\alpha>-g^2/(72)$, solitons indeed exist, and they pick up the $\tilde g_-$ branch.  In other words, effective cosmological constant of the soliton solutions are given by $\Lambda_{\rm eff}^-$.  Note that such AdS solitons are rather unusual in Einstein gravity.  As mentioned earlier, the embedding of this specific solution in Einstein gravity violates the null-energy condition.

\subsection{Extremal black holes with $T=\lambda$}

Exact solutions can be found for $f(T)$ theories of type (\ref{quadraticF}).  Let the parameter $\lambda=-6g^2$. The equations of motion are all satisfied provided that
\be
\fft{h'}{h} = \frac{-f+2 \sqrt{f}+3 g^2 r^2-1}{\left(f-\sqrt{f}\right) r}\,.
\ee
A solution of this is a black hole with the metric
\bea
ds^2 &=& - f dt^2 + \fft{dr^2}{f} + r^2 (d\theta^2 + \sin^2\theta\, d\Omega^2)\,,\nn\\
f &=& \Big(\sqrt{\fft{r_0}{r} +
\fft{g^2(r^3-r_0^3)}{r}}-1\Big)^2\,.
\eea
This describes an extremal black hole that are asymptotic to local AdS with the horizon located at $r=r_0$.

\section{AdS and Lifshitz planar black holes}
\label{adslifs}

In this section, we construct asymptotic AdS (or Lifshitz) planar black holes.  In this section, we consider the diagonal vielbein ansatz with toroidal symmetry, discussed in subsection \ref{tosy}.

\subsection{AdS planar black holes}
\label{adsplanar}

Making a coordinate gauge choice $A=h$ and $B=1/f$ and $\rho=r$, such that the metric becomes
\be
ds^2=-h dt^2 + \fft{dr^2}{f} + r^2 (dx^2 + dy^2)\,.
\ee
(Again, there should be no confusion between the metric function $f(r)=1/g_{rr}$ and the $f(T)$ function of the theory.) We find that the Schwarzschild AdS planar black hole is a solution, namely
\be
h(r)=f(r)=g^2\, r^2 - \fft{2M}{r}\,,\qquad \hbox{with}\qquad \Lambda_{\rm eff} = -3 g^2=\fft{f(T)}{4 f_T}\,.
\ee
Here $g=1/\ell$, the inverse of the radius of the AdS spacetime.  Note that for this AdS black hole, the torsion is a constant, given by
\be
T=-6g^2\,,
\ee
Thus for $f(T)$ gravity of the form (\ref{FT2}), the effective cosmological constant $\Lambda_{\rm eff}=-3g^2$ is related to $\alpha$ and the bare cosmological constant $\Lambda_0$ by
\be
\Lambda_0 = \Lambda_{\rm eff} - \ft12(2n-1)\alpha  (2\Lambda_{\rm eff})^n\,.
\ee

\subsection{Lifshitz black holes}

When $f(T)$ is given by (\ref{quadraticF}), equations can be all solved by $T=\lambda$.  For the toroidal-symmetric ansatz, this implies that the metric functions $(h,f)$ are related by
\be
f=-\fft{\lambda r^2 h}{2(h + r h')}\,.
\ee
The solutions are in general black holes, since $(h,f)$ can have common zeros at some radius $r=r_0$.  One interesting class of solution is the Lifshitz black hole
\bea
ds^2 &=& \ell^2\Big(-r^{2z} \hat f dt^2 + \fft{dr^2}{r^2 \hat f} + r^2 (dx^2 + dy^2)\Big)\,,\nn\\
f &=& 1- \fft{M}{r^{2z+1}}\,,\qquad \lambda=-\fft{2(2z+1)}{\ell^2}\,.
\eea
The solution reduces to the Schwarzschild AdS planar black hole when $z=1$. Analogous solutions was constructed in certain $f(R)$ gravity \cite{Cai:2009ac}.

\section{Charged black holes}

In this section, we consider $f(T)$ gravity coupled to a Maxwell field $A_\mu$ with
\be
{\cal L}=-\ft14 F^{\mu\nu} F_{\mu\nu}\,,\qquad F_{\mu\nu}=\partial_\mu A_\nu - \partial_\nu A_\mu\,.
\ee
The $f(T)$ theories that we consider here are (\ref{FT1}) and  (\ref{FT2}) with $n=2$.  Our solutions differ from those constructed in the UV approximation of the $f(T)$ theories \cite{Rodrigues:2013ifa}.

\subsection{Asymptotically-AdS solutions}

In this subsection, we use the toroidal-symmetric ansatz discussed in subsection \ref{tosy}, with the following coordinate gauge
\be
ds^2 = - h dt^2 + \fft{dr^2}{f} + r^2 (dx^2 + dy^2)\,,\qquad A=\phi dt\,.
\ee
The Maxwell equation implies that $\phi'=\sqrt{h/f}\, Q/r^2$.  The general exact solution is hard to come by, and we shall present only some special limit cases.

\subsubsection{Small $Q$ solutions}

As was seen in section \ref{adsplanar}, the AdS planar black hole is a solution of $f(T)$ gravity. Here we consider solutions with small electric charge $Q$. Up to and including the order of $Q^2$, the ansatz is
\be
h=\Big(g^2 r^2 - \fft{2M}{r}\Big) \Big (1 + Q^2 \tilde h + {\cal O}(Q^4)\Big)\,,\qquad
f=\Big(g^2 r^2 - \fft{2M}{r}\Big) \Big (1 + Q^2 \tilde f + {\cal O}(Q^4)\Big)\,,
\ee
with $\Lambda_0 = - 3g^2(1 + 18 \alpha g^2)$.
We find
\bea
\tilde h &=& -\frac{1}{g^2(1 + 12 \alpha g^2)  r_0 r\left(r^2+r r_0+r_0^2\right)}\,,\nn\\
\tilde f &=& \Big(1 - \fft{8\alpha g^2 r_0 (r^2 + r_0 r + r_0^2)}{(1 + 36\alpha g^2) r^3}\Big) \tilde h\,.
\eea
where $r_0$ is the horizon of the neutral black hole, related to $M$ as $M=\ft12 g^2r_0^3$.  We have chosen the integration constant such that the functions $\tilde h$ and $\tilde f$ are well behaved on and outside of the horizon, which remains to be at $r=r_0$.  Thus for
sufficiently small $Q$, the solution is a good approximation for electrically-charged black holes.

\subsubsection{Small-$\alpha$ solutions}

When $\alpha=0$, the theory is simply Einstein gravity coupled to the Maxwell field and a cosmological constant.  It hence admits the standard RN-AdS black hole solution, with
\be
h=f=f_0\equiv g^2 r^2 -\fft{2M}{r} + \fft{Q^2}{r^2}\,,\qquad \Lambda_0=-3g^2\,.
\ee
At the linear order of small $\alpha$, we find the solution of the $f(T)$ theory  is
\bea
h &=& f_0\, \Big(1 -18\alpha g^2 + \fft{\alpha}{5 r^6 f_0} \big(5 c r^5 - 2 (Q^4 + 30 g^2 Q^2 r^4 + 45 g^4 r^8)\big)+ {\cal O}(\alpha^2)\Big)\,,\nn\\
f &=& f_0\, \Big(1 + \fft{8\alpha Q^2}{r^4} + \fft{\alpha}{5 r^6 f_0} \big(5 c r^5 - 2 (Q^4 + 30 g^2 Q^2 r^4 + 45 g^4 r^8)\big)+  {\cal O}(\alpha^2)\Big)\,.
\eea
Here $c$ is an integration constant.  In order for the $\alpha$ contributions to be a small, we must can cancel the divergence arising from $f_0=0$ at the horizon $r=r_0$.  This can be indeed achieved if we set
\be
c=\fft{2(Q^4+30g^2 r_0^4 Q^2 + 45 g^4 r_0^8)}{5r_0^5}\,.
\ee
Then the solution is a good approximation of charged AdS planar black hole of the $f(T)$ theory for small $\alpha$.

\subsection{Asymptotically-flat solutions}

For asymptotically-flat solutions, we consider the ansatz discussed in subsection \ref{vielansatz1}, and make a coordinate gauge such that the metric takes the form of (\ref{hfmet1}).  We shall only study the $f(T)$ theory (\ref{FT1}) with $n=2$.  The exact solutions are also hard to come by and we shall consider various limiting cases.

\subsubsection{Small $\alpha$ solutions}

When $\alpha=0$, the theory becomes the standard Einstein-Maxwell theory and the corresponding RN black hole is given by
\be
h=f=f_0 \equiv 1 - \fft{2M}{r} + \fft{Q^2}{r^2}\,.
\ee
For the $f(T)$ gravity, we consider solutions of small $\alpha$.  At the linear order, the ansatz is given by
\be
h=f_0\, (1 + \alpha \tilde h)\,,\qquad f=f_0\, (1 + \alpha \tilde f)\,.
\ee
We find $\tilde f$ can be solved exactly, given by
\bea
\tilde f &=& \frac{15 Q^2 r^3 (r (c_1 r-40)+48 M)-120 M^2 r^4-80 Q^4 r (3 M+r)+114 Q^6}{15 Q^2 r^6f_0}\nn\\
&&-\frac{4 \left(-15 M^3 r^3-5 M^2 Q^2 r^2+10 M Q^4 r+25 M Q^2 r^3+6 Q^6-21 Q^4 r^2\right)}{3 Q^4 r^4 \sqrt{f_0}}\cr
&&-\frac{8 r \sqrt{M^2-Q^2} \left(2M^2-Q^2\right)}{Q^4 r^2f_0}
{\rm arctanh}\left(\frac{r-M}{\sqrt{M^2-Q^2}}\right)\cr
&& + \fft{4(M^2-Q^2)r}{Q^5r^2f_0} \Big(
5 \left(M^2-Q^2\right) \log \left(Qr \sqrt{f_0}-M r+Q^2\right)\cr
&&+\left(-5 M^2+4 M Q+5 Q^2\right) \log r-2 M Q \log \left(r^2f_0\right)\Big)\,.
\eea
The function $\tilde h$ can be solved up to a quadrature, given by
\bea
\tilde h'&=& \fft{1}{15 Q^2 r^9 f_0^2}\Big(-15 Q^2 r^5 \left(16 M-c_1 Q^2\right)-15 c_1 Q^2 r^7+120 r^6 \left(M^2+3 Q^2\right)\cr
&&-80 Q^2 r^4 \left(3 M^2+5 Q^2\right)-60 M Q^6 r+480 M Q^4 r^3+24 Q^8-44 Q^6 r^2\Big)\cr
&&+\fft{4(r^2-Q^2)}{3Q^4 r^7 f_0^{3/2}}\Big(-5 M r^3 \left(3 M^2-5 Q^2\right)-Q^2 r^2 \left(5 M^2+9 Q^2\right)-2 M Q^4 r+6 Q^6\Big)\cr
&& \fft{4(M^2-Q^2)(r^2-Q^2)}{Q^5 r^4 f_0^2} \Big(5 \left(M^2-Q^2\right) \left(\log r-\log \left( Q r \sqrt{f_0}-M r+Q^2\right)\right)\cr
&&+2 M Q \log f_0\Big) + \fft{8\sqrt{M^2-Q^2}(2M^2-Q^2)(r^2-Q^2)}{Q^4 r^4 f_0^2}
{\rm arctanh} \fft{r-M}{\sqrt{M^2-Q^2}}\,.
\eea
Here $c_1$ is the non-trivial integration constant. The solution becomes much simpler in the extremal limit, where $f_0=(1-Q/r)^2$. In this case, we have
\bea
\tilde h &=& -\frac{2 \left(3 Q^3+26 Q^2 r+69 Q r^2+192 r^3\right)}{15 Q r^4}
+\frac{5 c_1 Q-32}{5 Q (r-Q)}\nn\\
&&\qquad+\frac{5c_1 Q-2}{5 (r-Q)^2} - \fft{32}{Q^2}\log
\big(1 - \fft{Q}{r}\big)\,,\nn\\
\tilde f &=& \fft{2Q^3(19Q-20r)}{5(r-Q)^2 r^4} + \fft{r c_1}{(r-Q)^2}\,.
\eea
In both non-extremal or extremal cases, the integration constants $c_1$ is not sufficient to remove all the divergences of $(\tilde h,\tilde f)$ when $f_0=0$.  Thus the solution is only a good approximation away from the horizon.

\subsubsection{General asymptotic behavior}

We shall not attempt to find exact solutions for the generic $\alpha$ parameter.  Instead we shall only present the asymptotic behavior. It is easy to verify that the leading falloffs of the solutions at large $r$ are those of the RN black holes, with $\alpha$ contributions arising at the $1/r^5$ order:
\bea
h &=& 1 - \fft{2M}{r} + \fft{Q^2}{r^2} -\frac{8 \alpha  M^3}{5 r^5}+\frac{2\alpha  \left(30 M^2 Q^2-29 M^4\right)}{15r^6}\cr
&&\qquad\qquad-\frac{2 \alpha  M \left(405 M^4-586 M^2 Q^2+165 Q^4\right)}{105 r^7} + \cdots\,,\nn\\
f &=& 1-\frac{2 M}{r}+\frac{Q^2}{r^2} -\frac{8 \alpha  M^3}{r^5}+\frac{2\alpha  \left(50 M^2 Q^2-31 M^4\right)}{5r^6}\nn\\
&&\qquad\qquad -\frac{2 \alpha  M \left(29 M^4-50 M^2 Q^2+21 Q^4\right)}{3 r^7} + \cdots\,.
\eea
This general behavior indicates that the charged solutions carry the characteristics of the RN black hole at large $r$, and the difference between $f(T)$ theories and Einstein gravity are negligible in the infra-red region.

\section{Conclusions}

In this paper, we constructed classes of static solutions with spherical and toroidal isometries in $f(T)$ gravities, which are generalizations to the teleparallel equivalent of general relativity, analogous to the $f(R)$ generalization of the Einstein-Hilbert action. The fundamental fields of $f(T)$ gravities are the vielbein $e^a_\mu$ for a given metric; therefore, there are additional six new fields of the $SO(1,3)$ Lorentz group. This makes it much harder to classify the solution space. In this paper, we restricted our attention and regarded $f(T)$ gravities as some modified theories that deviate from Einstein gravity.  In other words, we focused on the $f(T)$ theory of the form (\ref{FT1}), which reduces to the teleparallel equivalent of Einstein gravity when the parameter $\alpha$ vanishes. In the large part of the paper, we also limited our interest in solutions that are asymptotic to the most symmetric vacuum, where the vielbein $e^{a}_\mu$ form an identity matrix, giving rise to Minkowski spacetime in Cartesian coordinates. Since $f(T)$ theories are invariant under general coordinate transformation, we could thus obtain the corresponding spherically symmetric ansatz.

We found that for the $f(T)$ theories (\ref{FT1}), the leading falloff associated with the $\alpha$ term is $1/r^{4n-3}$, which is much faster than the mass term of the Schwarzschild black hole.  In fact it is even much faster than the $Q^2/r^2$ in the RN black hole.  The results demonstrate that in the infra-red long wavelength region, the deviation of $f(T)$ gravities from Einstein gravity is small and they can easily pass the solar-system tests.  In the middle, however, numerical analysis indicates that there are naked time-like branch-cut singularities.

In the limit of large $\alpha$, the equations of motion can be solved by a single scalar equation $T=0$.  We found that extremal black holes could emerge.  A more intriguing class of solutions are perhaps the dark wormholes.  For any constant time slice of such a metric, the spatial section is analogous to that of an Ellis wormhole, but $-g_{tt}$ runs from 0 to 1 as a signal travels from one asymptotic region to the other, causing an arbitrarily large red shift.  As we commented in subsection \ref{darkworm}, such dark wormholes could not arise in Einstein gravity with proper minimally-coupled matter, but they can also emerge in some non-minimally coupled Einstein-vector theory.  The fact that such different theories give rise to these dark wormholes suggests that such a spacetime configuration deserves further investigation.

By introducing a bare cosmological constant $\Lambda_0$ to (\ref{FT1}), we obtained the smooth soliton configuration whose metric in the middle $r=0$ is like AdS in global coordinates of $\Lambda_0$, but flows to asymptotic locally AdS spacetimes with $\Lambda_{\rm eff}^-$ given in (\ref{lameffpm}). We also embedded the AdS planar and Lifshitz back holes in $f(T)$ gravities; furthermore, we coupled Maxwell fields to the $f(T)$ theories and constructed electrically-charged solutions.

\section*{Acknowledgement}

We are grateful to Shou-Long Li for proofreading the manuscript. The work was supported in part by NSFC grants 11175269, 11235003 and 11475024.

\end{document}